\documentstyle[rmp,aps,epsf,amsfonts]{revtex}
\begin{document}
\hyphenation{in-var-i-ant}
\draft
\twocolumn[\hsize\textwidth\columnwidth\hsize\csname@twocolumnfalse%
\endcsname
\title{Molecular Chirality and Chiral Parameters}

\author{A. B. Harris,\footnote{Electronic address:
harris@harris.physics.upenn.edu}
Randall D. Kamien,\footnote{Electronic address: kamien@physics.upenn.edu}
and T. C. Lubensky\footnote{Electronic address: tom@physics.upenn.edu}}
\address{Department of Physics and Astronomy,
University of Pennsylvania, Philadelphia, PA 19104}
\date{18 January 1999; revised 6 May 1999}
\maketitle
\begin{abstract}
The fundamental issues of symmetry related to chirality are discussed
and applied to simple situations relevant to liquid crystals.
We show that any chiral measure of a geometric object is
a pseudoscalar (invariant under proper rotations but changing
sign under improper rotations) and must involve three-point
correlations which only come into play when the molecule has
at least four atoms.  In general, a molecule is characterized by
an infinite set of chiral parameters.  We illustrate the fact that
these  parameters can have differing signs and can vanish at
different points as a molecule is continuously deformed into
its mirror image.  From this it is concluded that handedness
is not an absolute concept but depends on the property being
observed.  Within a simplified model of classical interactions,
we identify the chiral parameter of the constituent molecules
which determines the macroscopic pitch of cholesterics.
\hfill\break
\centerline{(18 January 1999; revised 6 May 1999)}
\end{abstract}
]

\tableofcontents
\section{Introduction}
Since the birth of stereochemistry 150 years ago with Pasteur's discovery
of handedness in molecules (Pasteur, 1848) interest in chiral molecules has
continued unabated.  The term chirality was first coined by Lord Kelvin
(Thomson,
1893) :

\begin{quote}
``I call any geometrical figure, or group of points,
{\sl chiral}, and
say it has chirality if its image in a plane mirror, ideally
realized, cannot be brought to coincide with itself.''
\end{quote}
Chirality permeates the entire fabric of the biological world.  Indeed,
life as we know it could not exist without chirality.
The function of fundamental components of the cell, like actin, myosin,
proteins, and lipids, relies upon their being chiral.  The
handedness of a molecule can affect its odor, potency, and toxicity.
Thus the synthesis of a single enantiomer of a compound is crucial for
the delivery of safe and effective pharmaceuticals and food additives.

Since chirality is the absence of inversion symmetry, a structure is either
chiral or it is not.  However, just as the degree of order of a
ferromagnet, which is either ordered or not ordered, can be quantified, so
the chirality of a structure can also be. A major theme of this article is
the development of quantitative measures of chirality and its impact on
physically measurable properties of materials.  There is an agreed
upon convention to answer
the ``yes"-``no" question of whether a molecule is chiral or not by
identifying chiral carbons (to be discussed in more detail below) to which
a handedness can be assigned via the Cahn-Ingold-Prelog rule (McMurry, 1992),
which orders the chemical groups attached to the carbon according to their
molecular weight.  The identification of the handedness in this way,
however, gives no indication of the magnitude, or even the sign of the
optical rotatory power this molecule will exhibit in solution.
Indeed the magnitude and sign of the rotatory
power depends on wavelength (though it is usually quoted in handbooks for
the D-line of Na).  Rotatory power provides only one
quantitative measure of chirality, which could be used, for instance
to assess whether one substance is {\sl more} or {\sl less} chiral than
another.
Similarly
when a liquid crystal forms a cholesteric it is obviously
chiral.  It can be more or less chiral depending on whether the pitch is
shorter or longer, respectively.
There are many other such quantitative measures.  In this paper,
we will define quantitative indices of chirality based on molecular
geometry and show how they enter the determination of a particular
observable, the pitch of a cholesteric liquid crystal.

\begin{figure}
\centerline{\epsfbox{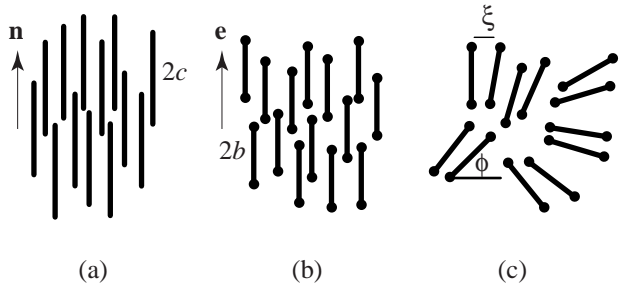}}
\vskip15pt
\caption{Schematic representation of (a) a nematic
liquid crystal in which
long molecular axes align on average along a spatially uniform director ${\bf
n}$.  The nematogens of this phase can either be uniaxial (like Fig.
{\protect\ref{biaxialfig}}a) or biaxial (like {\protect\ref{biaxialfig}}b).  In
the latter case, the long axis is the $c$ axis of length $2c$. (b) Schematic
representation of the plane perpendicular to ${\bf n}$ in a biaxial nematic.
The $b$ axes of nematogens align on average along ${\bf e}$ perpendicular to
${\bf n}$.  (c) Schematic representation of the plane perpendicular to ${\bf
n}$ in a uniaxial nematic composed of biaxial molecules.  There is no
long-range biaxial order but there are short-range orientational correlations
that persist out to a correlation length $\xi$.  The angle $\phi$ measures
the orientation of the the ``biaxial''-axis of each molecule with respect to
the $x$-axis.}
\label{nem_chol}
\end{figure}

\begin{figure}
\centerline{\epsfbox{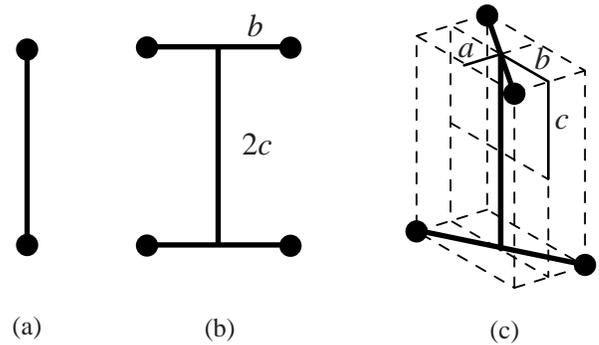}}
\vskip15pt
\caption{Representations of (a) Linear, (b) biaxial
planar, and (c) chiral molecules.
The molecule in (c) is a twisted ``H'' obtained from (b) by twisting about the
long molecular
axis (c-axis). It is both nonuniaxial and nonplanar as required for a
chiral molecule.}
\label{biaxialfig}
\end{figure}

Microscopic chiral constituents have a profound effect on
the macroscopic structures they form, striking
examples of which are common in liquid crystals (de Gennes and Prost, 1993).
The simplest liquid
crystalline phase is the nematic phase, characterized by long-range uniaxial
orientational order of anisotropic molecules called nematogens,
as shown in Fig.\ \ref{nem_chol}a.  The centers of mass
of the constituent molecules
are distributed homogeneously as in an isotropic fluid, but one of their
anisotropy axes aligns, on average, along a common unit vector ${\bf n}$ called
the director.  Strongly biaxial molecules (Fig.\ \ref{biaxialfig}) can in
principle condense into a biaxial rather than a uniaxial nematic phase with
long-range biaxial orientational order (de Gennes and Prost, 1993).
In this phase, one molecular
axis aligns along ${\bf n}$, and a second orthogonal axis aligns on average
along
a second vector ${\bf e}$ perpendicular to ${\bf n}$ as shown schematically in
Fig.\ \ref{nem_chol}b. Biaxial molecules can also condense into a uniaxial
nematic phase with short-ranged biaxial correlations rather than long-range
biaxial order as depicted in Fig.\ \ref{nem_chol}c.

\begin{figure}
\centerline{\epsfbox{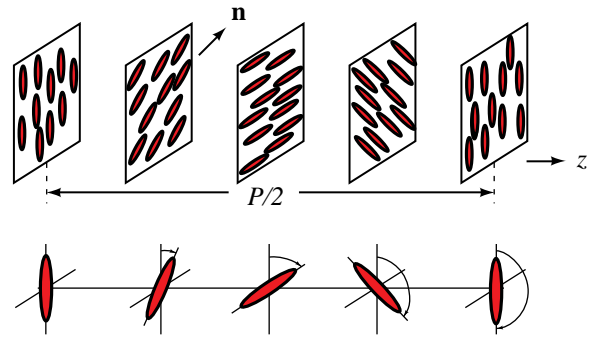}}
\vskip15pt
\caption{Schematic representation of a cholesteric
liquid crystal, showing the
helical twisting pattern of the local director ${\bf n}$ along the pitch axis
($z$ in this case).  The director rotates by $\pi$ in half a pitch $P/2$.}
\label{cholesteric}
\end{figure}

If these nematogens are chiral or if chiral
molecules
are added to an achiral uniaxial nematic, the director $\bf n$ will twist
creating the
simplest twisted phase: the cholesteric or twisted nematic
phase, the first liquid-crystalline phase to be
discovered (Reinitzer, 1888). This phase is
depicted schematically in Fig.\ \ref{cholesteric}.  The director at position
${\bf x} = (x,y,z)$ rotates in a helical fashion:
\begin{equation}
\label{neq}
{\bf n} ( {\bf x} ) = (\cos q z, \sin q z, 0) \ .
\label{twistn}
\end{equation}
In equilibrium the twist wavenumber $q$ assumes a preferred value $q_0$,
which corresponds to a pitch $P \equiv 2 \pi / q_0$.  Typically,
the pitch can vary from hundreds of nanometers
to many microns or more, depending on the system.  Cholesterics with
pitches on the order of $500$ nm Bragg scatter visible light and appear
iridescent.
If chirality is added to a
{\sl biaxial} nematic, a cholesteric structure similar to that shown in Fig.\
\ref{cholesteric} results, though the local molecular order will be strongly
biaxial. Other liquid-crystalline phases with macroscopic chiral
structure include
the blue phase, in which the director twists in all directions to produce a
three-dimensional periodic crystal, and the smectic-$C^*$ phase.
The dipole moments of the molecules in the latter phase become ordered as
a consequence of their chirality and make this phase ferroelectric.
Technologies based on these ferroelectric liquid crystals show
great promise for fast-switching, high-resolution displays.


\begin{figure}
\centerline{\epsfbox{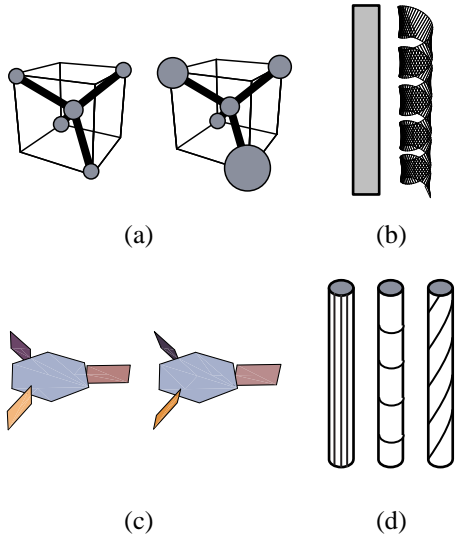}}
\vskip15pt
\caption{Examples of chiral structures created from
achiral ones: (a) left,
an achiral molecule in the shape of a tetrahedron with four equal masses
at its vertices; right, a similar chiral molecule
with four unequal masses at its vertices.
(b) left, an achiral planar sheet; right, a helix formed by twisting
a sheet about an cylinder. (c) left, an achiral
propeller with all blades perpendicular to the hexagonal core; right, a chiral
propeller with all blades rotated away from the normal to the hexagonal
plane.  (d) left and middle, cylinder with achiral
decorations; right, chiral cylinder with helical decorations.
}
\label{chiralmols}
\end{figure}

Models of chiral molecules are shown in Fig.\ \ref{chiralmols}.
Of particular importance in chemistry is the tetrahedrally coordinated
molecule, shown schematically in
Fig.\ \ref{chiralmols}a, consisting of a
central carbon atom with each of its four bonds connected to a
different chemical unit. If any two of these chemical units are equivalent then
the molecule has a mirror plane and is not chiral.  If all the chemical
units are different then the molecule is chiral and the central carbon atom is
referred to as
a chiral center.  More complex molecules may have many such chiral centers.
If all atoms of a molecule lie in a single plane, that plane is a mirror
plane, and the molecule is achiral; therefore chiral molecules must
be three-dimensional.   The converse does not hold:
not all three-dimensional molecules are chiral.  For instance, structures
that have continuous rotational symmetry about an axis (${\bf C}_\infty$)
are not chiral.
The simplest nonuniaxial structures have second-rank mass-moment tensors
with three inequivalent principal axes and are biaxial.
Fig.\ \ref{biaxialfig} shows model
linear, biaxial planar, and chiral molecules.  The twisted ``H" in Fig.\
\ref{biaxialfig}c is {\sl both} biaxial and chiral.

In spite of its practical importance, there is no universal quantitative
description
of molecular chirality nor is there an accepted procedure for
identifying the chiral part of an intermolecular interaction.  As a
result, only recently has real progress been made in addressing
straightforward and apparently simple issues such as the relation
between the cholesteric pitch and molecular geometry.  Ideally one
would like to introduce a parameter that measures the chiral
strength or degree of chirality of a given molecule.  A
non-vanishing value of this parameter, which we will refer to as a chiral
strength parameter or simply a chiral parameter, would distinguish a chiral
molecule from an achiral one, just as the dipole moment distinguishes a polar
molecule from a non-polar one.
In addition, such a chiral parameter would play a crucial role in
determining macroscopic chiral properties, such as the optical
rotary power, the wavevector for cholesteric ordering, and other
macroscopic chiral indices.
Unfortunately, there appears to be no such simple description of
chirality and chiral parameters.  As we will show, just as
a charge distribution can be described by an infinite hierarchy of multipole
moments,  so a chiral molecule can be described by an infinite hierarchy of
chiral parameters. If any one of these parameters is nonzero,
then the molecule is chiral.
Also, since different macroscopic properties will, in general, depend
on different microscopic chiral parameters, we do not expect strong
correlations between the various macroscopic manifestations of
chirality.

\begin{figure}
\centerline{\epsfbox{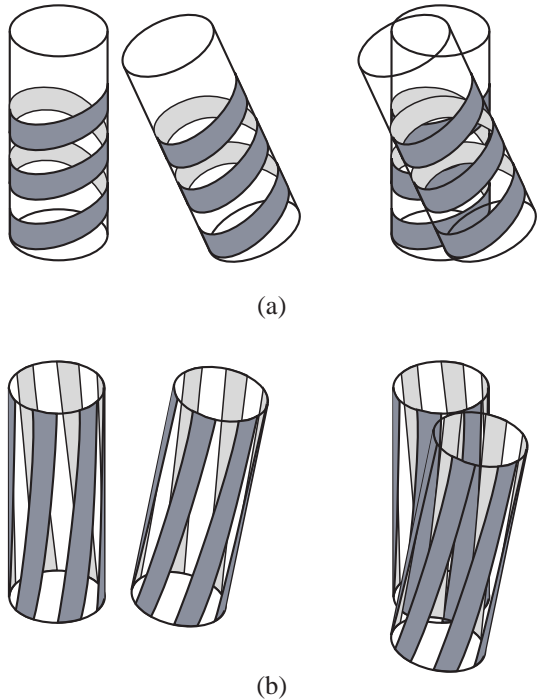}}
\vskip15pt
\caption{This figure shows how steric interactions
between two chiral
molecules produce a net relative rotation of their long axes.
The ``barber pole'' stripes on the model cylindrical molecules represent
protrusions such as are found on a screw.  If the protrusions on
one molecule (the shaded regions) are forced to fit ``hand-in-glove'' into the
grooves on another (the unshaded regions) then, as we show on the right, when
one molecule is placed
upon another,
the protrusions on one molecule align with the grooves on the
other.  The result is that the long axes of the two
molecules acquire a relative twist determined by the pitch of the
stripes. In (a) we show molecules with a tight right-handed pitch as
determined by the geometrical right-hand rule.  The relative twist of two
neighboring molecules is right-handed according to the right-hand rule.
(b) shows molecules with a weak right-handed pitch.  The relative twist of
neighboring molecules is now left-handed.  These examples show that the
``handedness" of individual molecules does not determine the handedness
of the collective structure.}
\label{straley_screws}
\end{figure}

Even if one were equipped with a complete understanding of the nature of
interactions
between chiral molecules, the calculation of macroscopic parameters
like the cholesteric pitch is not completely straightforward.  An argument due
to Straley (1974) and illustrated in Fig.\ \ref{straley_screws}
makes it clear why molecular chirality causes macroscopic rotation:
when two screw-like molecules are brought close
together, their grooves interlock to produce a finite rotation angle
$\Delta \theta$ between long molecular axes.  A simple estimate of the pitch
based on this picture is $P \approx (2 \pi / \Delta \theta ) l$ where $l$ is
the molecular diameter.  Taking a rough estimate of $\Delta\theta\approx
10^\circ - 20^\circ$ and a molecular size of roughly $1$ nm, one finds
that $P \approx 10$ nm, two or three orders of magnitude smaller
than typical pitches. Indeed some chiral systems (Fraden, 1995) are labeled
nematic rather than cholesteric, presumably because their pitches are too
long to be measured experimentally.
Thus a quantitatively correct calculation of the cholesteric pitch
cannot be
based solely on molecular parameters and presents a challenge to theorists.
As we demonstrated
previously (Harris, Kamien, and Lubensky, 1997), for central-force
or steric models, $q_0$ vanishes (infinite pitch) unless biaxial correlations
between
the orientations of adjacent molecules (such as are illustrated in Fig.\
\ref{nem_chol}) are taken into account.  Quantum interactions, however,
do not require such correlations
and hence can give a nonzero value of $q_0$ even within mean field theory.
%In particular, we showed that
%a proper calculation within mean-field theory (which neglects these
%correlations) {\sl must} give $q_0=0$.

The purpose of this paper is to present recent progress both in
quantitatively characterizing molecular chirality and in calculating
the cholesteric pitch from microscopic interactions.  In Section II we
start by making some na\"\i ve qualitative comments about the nature of
chiral symmetry.  The central idea is that
an achiral object has higher symmetry than a chiral one.
We will develop a systematic procedure for generating a countably infinite
set of chiral molecular parameters that
all vanish when the molecule is achiral.
Next, in Section III, to illustrate our chiral parameters
we consider a topological ``rubber
glove'', a chiral structure that can be converted to its mirror image via
distortions through a continuum of intermediate states all of which are
chiral.
This demonstration shows clearly that handedness is not an absolute
concept, but depends on the property under consideration.
In Section IV we describe a calculation of the cholesteric pitch from a
classical model of central forces between atoms.  This calculation shows that
biaxial correlations play a critical role in determining
the pitch $P$. In fact, if
these biaxial correlations do not exist, each molecule rotates freely and
appears, on average, uniaxial and thus achiral.
In accord with our discussions of chiral parameters in Section II,
we expect
that other macroscopic chiral response functions, such as the rotatory power,
will depend on other
chiral structure parameters.  Indeed, it is likely
that an understanding of many such indices of
molecular chirality is required to interpret the dramatic frequency dependence
of these susceptibilities.

\section{Chiral Parameters}

As mentioned in the introduction, one expects the chiral interaction
between molecules to involve parameters
characterizing the chiral strengths of the molecules.  However, there
is no obvious precise quantitative formulation of parameters that
characterize the degree to which a given molecule is chiral.
Only a handful of chiral strength parameters have been proposed.  For
instance, Osipov {\sl et al.} (1994) developed a measure of molecular
chirality by considering the symmetry of response functions describing the
electromagnetic behavior of chiral molecules.
We also introduced a chiral strength parameter
in a previous calculation (Harris, {\sl et. al}, 1997) of the chiral
wavevector $q_0$.
We will review this calculation of $q_0$
in Section IV.  Both of the above chiral parameters are non-local (in a
sense to be made more precise later) -- a feature which we shall see is
generic.   In the following
we give a more systematic discussion of the structure of such chiral
strength parameters.

\subsection{Chirality is the Absence of Symmetry}

A preliminary remark is that
``chiral symmetry'' is actually an absence of symmetry, {\sl i.e.},
the absence of symmetry under improper rotations.
Thus a chiral object has {\sl lower} symmetry than an achiral object which is
invariant under chiral operations.
This situation
contrasts with ``spherical symmetry'' which implies the
existence, rather than the nonexistence, of symmetry elements.
Accordingly, it is instructive to compare the way spherical symmetry
is destroyed when a sphere is distorted to the way achiral symmetry
is destroyed when an achiral object is chirally distorted.  To start,
we consider distortions of a sphere
centered at the origin.  Initially, the sphere of radius
$r_0$ is a surface described by
\begin{equation}
\label{rzero}
r = r_0 \ ,
\end{equation}
where $r$ is the radial coordinate from the origin.
When the sphere is distorted, the radial coordinate of its surface will depend
on
the usual angles $\theta$ and $\phi$ and can be expanded in
spherical harmonics:
\begin{equation}
\label{rnm}
r = r_0 + \sum_{n=1}^\infty \sum_{m=-n}^{m=n} a_{nm}
Y_{nm} ( \theta , \phi ) \ ,
\end{equation}
where $a_{nm}=a_{n,-m}^* (-1)^m$.

Usually one characterizes distortions by the values of the
$a_{nm}$'s for the smallest value of $n$ for which one of these is
nonzero.  The first-rank tensor $a_{1m}$ describes translations of the
sphere, which do not alter the symmetry and which we ignore.  Thus, the
lowest-order distortions are characterized by $a_{2m}$,
which in a Cartesian representation is a symmetric, traceless,
second-rank tensor.  In general, a complete specification of the shape of an
aspherical surface requires the values of the infinite set of
$a_{nm}$.   Since the $a_{nm}$ mix with each other under rotation,
it is desirable to construct rotationally invariant measures of asphericity.
A useful class are the quantities
\begin{equation}
\label{esig}
\sigma_n \equiv \sum_m a_{nm}^2 ,
\end{equation}
which provide rotationally invariant characterizations of the magnitude of
the asphericity associated with $n$th-rank tensor distortions.  It is
entirely possible for $\sigma_2$ to vanish while higher-order $\sigma_n$ do
not.  In this case the distortion is characterized by the lowest-order,
nonvanishing value of $\sigma_n$.

With the above discussion in mind, we consider chirality.
An object
can be described by the multipole moments of its density,
$\rho(r, \theta, \phi)$, namely
\begin{equation}
\label{edens}
\tilde \rho_{lmN}  \equiv  \int d{\bf r} \rho(r,\theta,\phi) r^N
Y_{lm}(\theta,\phi)\ ,
\end{equation}
where $\tilde \rho_{lmN}^* = (-1)^m \tilde \rho_{l,-m,N}$ since
$\rho(r,\theta,\phi)$ is real.
The moments $\tilde \rho_{lmN}$ for a given $N$ define a tensor
parameter that transforms under a ($2 l+1$)-dimensional representation
of the rotation group.  The alternative Cartesian representation in terms
of symmetric, traceless tensors $\tilde \rho_{lN}^{i_1 ...i_l}$ of rank $l$
is used extensively in treatments of liquid-crystalline order, and we will
employ
them when appropriate.
For a molecule consisting of point atoms, the density consists of
a sum of Dirac-delta functions locating each atom.  This provides a
natural framework to study interactions between molecules, in which
connection the central quantities are $\tilde \rho_{lmN}$.
The question we wish to address here is how
these moments, or appropriate functions of them, characterize chirality.
We start by discussing the analogs
of the parameters $\sigma_n$ in order to characterize the magnitude of
chirality.

\subsection{Construction of Pseudoscalars}

Bearing in mind that chirality requires the absence of
inversion symmetry, we propose to characterize the magnitude
of chirality by an infinite sequence of pseudoscalars.  First,
note that Lord Kelvin's definition may alternatively be stated as
``an object is achiral if there exists a rotation $\bbox{\Omega}$ such that
the object is invariant under the operation $\bbox{\Omega}{\bf S}_2$, where
${\bf S}_2$ is spatial inversion.''
Since spatial inversion is a mirror operation followed by a $\pi$ rotation
about an axis perpendicular to the plane of the mirror, this definition
of chirality is equivalent to Lord Kelvin's.  A scalar is invariant
under both rotations and inversion while a pseudoscalar is only
invariant under rotations -- it changes sign under inversion.  Thus any
pseudoscalar parameter $\psi_n$ that we construct from the multipole moments
of the density will necessarily vanish when the molecule is {\sl achiral}.
Furthermore, the degree of chirality can be characterized by the
magnitudes of the set of $\psi_n$ just as the degree of asphericity
was characterized by the various $\sigma_n$.

The construction of pseudoscalars can be done systematically by
considering the representation theory of the three-dimensional rotation
group $O(3)$.  This procedure amounts to nothing more than considering
the quantum-mechanical addition of angular momentum.
To each representation we will attach its transformation
properties under inversion ({\sl i.e.}, parity).
Pseudoscalars
will transform as 1-dimensional representations with odd parity.  Denoting
the $d$-dimensional representation with parity $p$ as ${\bf d}^p$, we
note that the rank-$l$ representations generated by the multipole expansion
are
\begin{equation}
\label{ereps}
{\bf 1}^+,{\bf 3}^-,{\bf 5}^+,\ldots,{\bf (2l+1)}^{(-)^l},\ldots .
\end{equation}
To construct a pseudoscalar we must form tensor products of
different representations.  While the resulting representations obey
the rules for addition of angular momentum, the parity of the new
representation is simply the product of the parities of the
two representations.  For instance, since two spin-1 states (with odd parity)
can be combined
to form a spin-2, spin-1 or spin-0 state (all with even parity), we have
\begin{equation}
\label{ett}
{\bf 3}^- \otimes{\bf 3}^- = {\bf 5}^+ \oplus{\bf 3}^+ \oplus{\bf 1}^+ \ .
\end{equation}
This gives us our first representation that is not a multipole representation:
${\bf 3}^+$ a
{\sl pseudovector}.  Forming the triple tensor product:
\begin{eqnarray}
\label{ettt}
{\bf 3}^-\otimes&{\bf 3}^-&\otimes{\bf 3}^- =
\nonumber\\
&&\quad{\bf 7}^-\oplus{\bf 5}^-\oplus{\bf 5}^-\oplus{\bf 3}^-\oplus{\bf 3}^-
\oplus {\bf 3}^-\oplus{\bf 1}^-  ,
\end{eqnarray}
we arrive at our first pseudoscalar ${\bf 1}^-$, which we recognize
as the
vector triple product ${\bf A}\cdot\left({\bf B}\times{\bf C}\right)$.
The above discussion suggests that any pseudoscalar must involve a
product of {\sl at least} three of the multipole moments $\tilde \rho_{lmN}$.
It is, in fact, always the case that a ${\bf 1}^-$ can only be constructed from
the tensor product of ${\bf d}^+\otimes{\bf d}^-$.  Since the
multipole moments do not include {\sl both} ${\bf d}^+$ and ${\bf d}^-$,
one must construct one of these via tensor products.  Thus any pseudoscalar
must involve a product of {\sl at least} three of the multipole moments.
This implies that a chiral parameter for a given object can be expressed as
an integral over at least three position vectors in that object, and in this
sense, chirality is a nonlocal property.

\subsubsection{Nonpolar Molecules}
In this subsection we confine our attention to the case in which vector
representations can be eliminated by proper choice of the center of the
molecule.  Specifically, in this case we do not allow molecules to have
a dipole moment.  (That case will be considered in the next subsection.)
To illustrate this theoretical discussion, let us look for the lowest
order (in powers of components of ${\bf r}$) pseudoscalar that can be
constructed from the simple mass-weighted distance moments
\begin{equation}
\label{emass}
\rho_{lm} = \sum_{\chi\in\scriptscriptstyle X}
\vert{\bf r}_\chi\vert^l Y_{lm} (\theta_\chi ,
\phi_\chi) \
\end{equation}
of a homoatomic molecule,
where the sum is over atoms $\chi$ in the molecule $X$.
Throughout, we will label molecules with capital Roman letters and their
constituent atoms
by Greek letters.  The center of mass of molecule $A$ will be ${\bf
R}_{\scriptscriptstyle A}$,
and each atom $\alpha$ will be displaced from there by ${\bf
r}_{{\scriptscriptstyle A}\alpha}$.
For simplicity
we focus only on $\rho_{lm}\equiv \tilde \rho_{lml}$.  Our discussion
could be embellished by considering $\tilde \rho_{lmN}$ for other values of
$N$.
If we measure the density relative to the center of mass, then
$\rho_{1m}=0$ for all $m$.  Thus the multipole
expansion only provides us with $d=5$ or larger dimensional representations.
Since we have restricted ourselves to a single tensor for each $d$-dimensional
representation, ${\bf 5}^+ \otimes {\bf 5}^+$ will not contain any
pseudotensors, and
the lowest-order pseudotensor we can construct is contained in
${\bf 5}^+ \otimes {\bf 7}^-$.  We could now try to construct a pseudoscalar by
contracting the resulting tensors with ${\bf 5}^+$, but, again because we are
considering only moments of $\rho_{lm}$, we would get zero just as we would get
zero for the triple product ${\bf A} \times {\bf B} \cdot {\bf A}=0$ in the
vector case. A nonzero
pseudoscalar only results when ${\bf 5}^+\otimes{\bf 7}^-$
is contracted with a tensor {\sl different} from the ${\bf 5}^+$
and the ${\bf 7}^-$.  Thus the lowest order pseudoscalar we seek is in
${\bf 5}^+ \otimes {\bf 7}^- \otimes {\bf 9}^+$.  In terms of
spherical harmonics we set
\begin{equation}
\label{epsiz}
\psi_0 \propto \sum_{mn} C(234;mn) \rho_{2m} \rho_{3n} \rho_{4,m+n}^* \ ,
\end{equation}
where $C(234;mn)$ are the appropriate Clebsch-Gordon coefficients.
It is convenient to choose the normalization so that in
the Cartesian representation, this is
\begin{eqnarray}
\label{epsizb}
\psi_0 = \rho_2^{ij} \rho_3^{klm} \epsilon_{ikp} \rho_4^{jlmp}\ .
\end{eqnarray}
Note the presence of the antisymmetric symbol $\epsilon_{ijk}$ in this
expression.  It is required to produce a scalar from two even-ranked and one
odd-ranked tensor.

We can calculate $\psi_0$ for the ``twisted H'' molecule, $M_1$, shown in Fig.\
2,
with four identical atoms at
\begin{equation}
M_1=\{(a,b,c),(a,-b,-c),(-a,b,-c),(-a,-b,c)\} \ .
\end{equation}
This molecule is chiral if $abc\ne 0$ and if $\vert a\vert\ne \vert b\vert\ne
\vert c\vert\ne \vert a\vert$.
Since $\psi_0$ is a rotational invariant, it may be evaluated in any
convenient coordinate basis.  We find
\begin{eqnarray}
\label{epsires}
\psi_0 = K_0 (a^2-b^2) (b^2-c^2) (c^2-a^2) abc \ ,
\end{eqnarray}
where $K_0$ is a numerical constant.  Note that $\psi_0$ does indeed
vanish when the parameters assume the special values for which
the molecule has the higher achiral symmetry. Moreover, since the mirror
image of $M_1$ may be obtained by exchanging any two of $a$, $b$ and $c$ or
by reversing any one of their signs, we see that under inversion
$\psi_0\rightarrow -\psi_0$, and it is indeed a pseudoscalar.  This
ninth-order
multinomial is the lowest-order expression constructed from $\rho_{lm}$,
which must vanish for an achiral object.
However, just as in the discussion of asphericity, it is possible
to consider a class of chiral molecules for which $\psi_0$ vanishes
but which requires an even higher-order multinomial to characterize
its chirality.  Consider a twelve atom molecule, $M_2$, obtained by taking
the atoms as in the ``twisted H'' together with the eight atoms
obtained by cyclic permutation, so that identical atoms are now at
\begin{eqnarray}
M_2&=\bigg\{&(a,b,c),(a,-b,-c),(-a,b,-c),(-a,-b,c),\nonumber\\
&&(b,c,a),(-b,-c,a),(b,-c,-a),(-b,c,-a),\\
&&     (c,a,b),(-c,a,-b),(-c,-a,b),(c,-a,-b)\bigg\}.\nonumber
\end{eqnarray}
To show that
$\psi_0$ vanishes for $M_2$, it suffices to verify that
$\rho_{2m}=0$ for all $m$.  (This result is most easily
verified using the Cartesian representation for $\rho_2^{ij}$.)
However,  except for the special values of the parameters
({\sl i.e.}, $a=0$, $b=0$, $c=0$, $\vert a\vert=\vert b\vert$, $\vert
b\vert=\vert c\vert$, or $\vert c\vert=\vert a\vert$),
this object is clearly still chiral since it is the union of three
identical chiral objects. To describe its chirality the lowest order
pseudoscalar constructed from the moments $\rho_{lm}$, is
\begin{eqnarray}
\label{pseqb}
\psi_1 = \sum_{\mu \nu} C(346;\mu \nu) \rho_{3\mu} \rho_{4\nu}
\rho_{6,\mu+\nu}^* \ ,
\end{eqnarray}
which we evaluate to be
\begin{eqnarray}
\label{eval}
\psi_1 &=& K_1 abc (a^2-b^2)(b^2-c^2)(c^2-a^2)\nonumber\\
&&\quad\quad\times(a^4 + b^4 + c^4
-4 a^2 b^2 - 4 b^2 c^2 - 4 c^2 a^2 ),
\end{eqnarray}
where $K_1$ is a numerical constant.  As was the case for $\psi_0$,
this chiral strength parameter vanishes when the ``twisted H'' is made to be
achiral.

It is clear that we can construct an
infinite sequence of chiral parameters from the $\rho_{lm}$
that vanish for achiral objects.  For
example, when $J$, $K$, and $L$ are all different integers whose sum is
odd, each of the quantities
\begin{equation}
\psi_{JKL} = \sum_{mn} C(JKL;mn) \rho_{Jm} \rho_{Kn} \rho_{L,m+n}^* \
\end{equation}
is a chiral parameter, which is nonzero only for chiral molecules.

If we consider different tensors of a given rank, we can construct
other sets of chiral parameters not encompassed by $\psi_{JKL}$.
For example, if there are two distinct second-rank tensors
$\gamma_2^{ij}$ and $\tau_2^{ij}$, then we can construct the
chiral parameter
\begin{eqnarray}
\label{eorda}
\psi_2 & = & \!  \gamma_{2}^{il} \tau_{2}^{jm}
\epsilon_{ijk} \rho_3^{klm} \nonumber \\ &\propto&
\!\sum_{mn} C(223;mn) \gamma_{2m} \tau_{2n} \rho^*_{3,m+n} .
\end{eqnarray}
When $\gamma_2^{ij} = \tau_2^{ij}$, $\psi_2$ vanishes because $\epsilon_{ijk}$
is antisymmetric in all indices.  The tensors $\gamma_2^{ij}$ could,
for example,
be constructed from
$\rho_{2m}$ and ${\tilde \rho}_{2mN}$ for $N\neq l$.  Alternatively,
two different tensors can
be constructed from the second-rank mass-moment tensor
\begin{equation}
\rho_2^{ij} = \sum_{\chi\in X} \left(r_{\chi}^i r_{\chi}^j - \case{1}{3}
r_{\chi}^2 \delta_{ij} \right) .
\end{equation}
To do this, we express $\rho_2^{ij}$
in the basis of its orthonormal principal axes emblazened on the molecule
${\bf e}_1$, ${\bf e}_2$, and ${\bf e}_3$,
where ${\bf e}_3= {\bf e}_1 \times {\bf e}_2$ is
associated with the largest magnitude eigenvalue of $\rho_2^{ij}$.
Then we decompose $\rho_2^{ij}$ into its uniaxial (${Q^{ij}}$) and
biaxial (${B^{ij}}$) components as
\begin{equation}
\rho_2^{ij} =S Q^{ij} + B\left(e_1^i e_1^j - e_2^i e_2^j\right)\equiv S Q^{ij}
+B^{ij} ,
\end{equation}
where
\begin{mathletters}
\label{SBQEQ}
\begin{eqnarray}
S&=&{3\over 2}\sum_{\chi}\left[\left({\bf r}_\chi \cdot {\bf e}_3\right)^2 -
{1\over 3}
{\bf r}_{\chi}^2\right] \\
B &=& {1\over 2}\sum_{\chi}\left[\left({\bf r}_{\chi} \cdot
{\bf e}_1 \right)^2 -
\left({\bf r}_{\chi} \cdot {\bf e}_2 \right)^2\right]
\end{eqnarray}
and
\begin{equation}
Q^{ij} = (e_3^i e_3^j - {1\over 3}  \delta_{ij} ) .
\end{equation}
\end{mathletters}
Note that $B^{ij}$, the biaxial part of $\rho_2^{ij}$,
vanishes when the molecule is uniaxial.
Setting $\gamma_{2}^{ij} = Q^{ij}$, $\tau_{2}^{ij} = B^{ij}$, and
$\rho_3^{ijk} \equiv S^{ijk}$, we obtain
\begin{equation}
\psi_2 = Q^{il} B^{jm}\epsilon_{ijk} S^{klm}
\label{psi2}
\end{equation}
as a chiral strength parameter, which plays a role in our calculation
of the cholesteric pitch $q_0$ to be presented in Section IV.
For the ``twisted H'' molecule, we have
\begin{eqnarray}
\label{PSIEQ}
\psi_2 =&& C_2 abc\left[ \vert a^2 + c^2 -2b^2 \vert +a^2 - c^2 \right]
\nonumber \\
&& {\rm for} \ \ \  a^2 \leq b^2 \leq c^2 \ ,
\end{eqnarray}
where $C_2$ is a numerical constant.
Expressions for $\psi_2$ in regimes other than $a^2 \leq b^2 \leq c^2$
can be obtained by suitably permuting variables.  One may verify
that when Eq. (\ref{PSIEQ}) is valid, $\psi_2$ does vanish when the
molecule is achiral, {\sl i.e.} when $a^2=b^2 <c^2$ or $a^2<b^2=c^2$

One can construct different second-rank tensors in other
ways.  For instance,
in some phenomenological intermolecular potentials, the
strength of the dispersion interaction between a pair of
atoms is estimated to scale with the product of their
atomic polarizabilities.  In that case, a polarizability-weighted
second-distance moment is generated by the multipole expansion
of the intermolecular potential.  In general,
a tensor with any weighting that is distinct from the mass weighting
can play the role of the additional second-rank
tensor needed to characterize chirality.
Such moments would have the form
\begin{equation}
\bar\rho_{lm} = \sum_{\chi\in\scriptscriptstyle X}
w_\chi Y_{lm} (\theta_\chi, \phi_\chi) \ ,
\end{equation}
where $w_\chi$ is a weight factor, which can differ from
the factor $r_\chi^l$ in Eq. (\ref{emass}).  For a molecule with $p$ atoms,
we can obviously have up to $p$ linearly independent second-rank tensors.  As
we mentioned, moments similar to these have been used in the study of optical
properties
of chiral systems by Osipov {\sl et. al} (1995).

\subsubsection{Ferroelectric Liquid Crystals}
There are cases in which one may invoke the vector representation, even
in liquid crystalline systems. A particularly interesting case
is that of ferroelectric liquid crystals (Meyer, {\sl et al.}, 1975).
These phases are composed of mesogens that have an electric dipole
moment.
Recall that the smectic-C liquid crystalline phase is a one-dimensional
layered structure with layer normal ${\bf N}$.  In each layer the nematic
orientation ${\bf n}$ is not parallel to ${\bf N}$.  Thus, one can
construct
the pseudovector ${\bf A} = ({\bf n}\cdot{\bf N})({\bf n}\times{\bf N})$.
It is clear that under parity ${\bf A}$ does not change sign and
that both the nematic (${\bf n}\rightarrow -{\bf n}$) and layer normal
(${\bf N}\rightarrow -{\bf N}$) inversion symmetries are preserved.
However,
if the molecules are chiral then, as we have seen, a nonvanishing pseudoscalar
$\psi$ may be constructed.  In this case ${\bf P} = \psi{\bf A}$ is a true
vector and can set an unambiguous alignment direction (perpendicular to ${\bf
n}$ and ${\bf N}$) for the molecular dipole moments.

In general, when electrostatic interactions are taken into account,
both signs of charge are present and there is a non-zero
dipole moment that no change of origin can eliminate.  In this case,
we can again construct two different second-rank tensors $\tau_2^{ij}$ and
$\gamma_2^{ij}$, and then
\begin{eqnarray}
\label{eordg}
\psi_f = \sum_{mn} C(122;mn) \rho_{1m} \tau_{2n} \gamma_{2,m+n}^*
\propto \rho_1^i \tau_2^{jk}\gamma_2^{lk}\epsilon_{ijl},
\end{eqnarray}
is a pseudoscalar which includes the (dipole) charge moment $\rho_{1m}$.
\begin{eqnarray}
\label{emoma}
\rho_1^i = \sum_{\chi\in\scriptscriptstyle X} q_\chi r^i_\chi .
\end{eqnarray}
In fact one could construct a pseudoscalar from a vector triple product of
three
noncoplanar vectors obtained by introducing three different weight
factors into the sum in Eq. \ (\ref{emoma}).

\section{``Rubber Glove'' Molecules}
We have argued that a quantitative characterization of
chirality does not rest on one parameter, but rather on an infinite
hierarchy of chiral moments.  However, one has a natural tendency to
associate a
specific handedness to a given
chiral molecule.   We will show that even the ``handedness'' of a molecule
depends on
the chiral property under consideration.  This is, in fact, familiar from
circular dichroism measurements: the difference in attenuation of
left- versus right-circularly polarized light changes sign as a
function of its wavelength.  Thus the handedness of an object is
really in the eye of the beholder.

\vbox{\begin{table}
\caption{\baselineskip=0.33truein Atoms and weightings $W$ and $W'$ for the
chiral
parameters $\psi_0$ and $\psi_0'$ of the ``twisted H'' molecule,
$M_1$.}
\label{T1}
\begin{tabular}{ccc}
Position ($\chi$) & $W(\chi)$ & $W'(\chi)$ \\
\hline
$(a,b,c)$ & 1 & $1 + \mu$ \\
$(a,-b,-c)$ & 1 & $1 + \mu$ \\
$(-a,b,-c)$ & 1 & $1 - \mu$ \\
$(-a,-b,c)$ & 1 & $1 - \mu$ \\
\end{tabular}
\end{table}}

To illustrate this idea, we consider a process in which a chiral molecule
is continuously deformed into its enantiomer or mirror-image molecule.
For the ``twisted H'' molecule, we could do this by continuously varying
the parameter $a$ until its final value becomes the negative of its
initial
value.  Obviously, when $a$ passes through zero the molecule is achiral
and one might be tempted to say that the plane $a=0$ in parameter space
separates regions of opposite handedness.  However, as we will show by
example, it is possible to continuously deform a chiral molecule into
its enantiomer {\sl without} passing through an achiral configuration.
(Here ``deformation" is used in its most general sense in which not
only the coordinates, but also the mass and other properties of atoms
are varied.)  The existence of such a continuous deformation is
incompatible with the existence of an intrinsic definition of
right or left handedness.  It is also clear that any single
measure of chirality will pass through zero at some point in the
process of deforming a molecule into its enantiomer.
However, a molecule is achiral {\sl only} if all of its chiral moments are
zero;
the vanishing of a single chiral moment alone does not make a molecule achiral.
We will illustrate explicitly that there exist paths of deformations
between enantiomers along which there is no point where all chiral moments
vanish.  Nevertheless, along this path every chiral
measure must and does vanish at some point.
A molecule that can be deformed in this
way is known as a topological rubber glove in analogy with a real rubber
glove -- it can be inverted one finger at a time thus always remaining chiral
(Walba, {\sl et al.}, 1995). In the context of our discussion we would
interpret this by saying that the eye automatically measures many indices of
chirality, and as each finger is inverted some indices may pass through zero
to change sign until finally all indices have changed sign.

We can see this explicitly by keeping track of more than
one chiral parameter as the ``twisted H'' is inverted continuously
into its enantiomer.  The two chiral parameters we will monitor
are $\psi_0$ and $\psi_0'$. $\psi_0$ is defined in
Eq. (\ref{epsiz}) and given explicitly in Eq. (\ref{epsires}).
$\psi_0'$ is also defined as in Eq.(\ref{epsiz}), except
that now $\rho_{lm}$ is replaced by $\rho_{lm}[W']$, where
\begin{equation}
\rho_{lm}[W] = \sum_{\chi\in\scriptscriptstyle X} W(\chi)
\vert{\bf r}_\chi-{\bf r}^0[W]\vert^l Y_{lm} (\theta_\chi , \phi_\chi) \ ,
\end{equation}
where $W(\chi)$ is a weighting function associated with the atom $\chi$
and ${\bf r}^0[W]$ is the
$W$-weighted center of the molecule, chosen so that $\rho_{1m}[W]$ vanishes.
Various weighting functions are shown in Table I.
Until now we have considered molecules composed of identical atoms, {\sl
i.e.} $W(\chi)\equiv 1$.   Of
course, the molecule need not have identical atoms, and, therefore, not all
properties of the atoms need be the same.  For instance, if all
the atoms are weighted equally,
$\rho_{lm}[W]$ would correspond to a purely geometric moment.
However, if we were to weight the atoms by their polarizabilities, then the
moments would be different.  We have defined $\psi_0$ to reflect geometric
properties and $\psi'_0$ to reflect others.
We could have instead introduced other weight functions
which reflect the valence, electronegativities, {\sl etc.}, of
the atoms.   Since different
properties are not perfectly correlated, they may require
different weight functions.
To construct the continuous
deformation between enantiomers, we only invoke ``twisted
H" molecules that have atoms with polarizabilities $1+\mu$ on sites \#1
and \#2 and $1-\mu$ on sites \#3 and \#4.
The point here is that the molecule is chiral if {\sl either}
$\psi_0$ or $\psi_0'$ is nonzero.

In the calculation of $\psi_0'$ all displacements ${\bf r}_\chi$
are evaluated relative to the ``center of polarizability'' of the molecule,
so that $\rho_{1m}'=0$ for all $m$.  We find that
\begin{eqnarray}
\psi_0'& =&K_0(1-\mu^2)^2 abc(b^2-c^2) \nonumber \\ &&
\ \ \times [(c^2-a^2)(a^2-b^2)-\case 9/7 \mu^2 a^4] \ .
\end{eqnarray}
When the molecule is tetrahedral ($a=b=c$), it is truly achiral
if any two of its atoms are identical.  Indeed in this case
$\psi_0$ and $\psi_0'$ (as well as all other chiral parameters)
vanish.

\begin{figure}
\epsfxsize=3.0truein
\centerline{\epsfbox{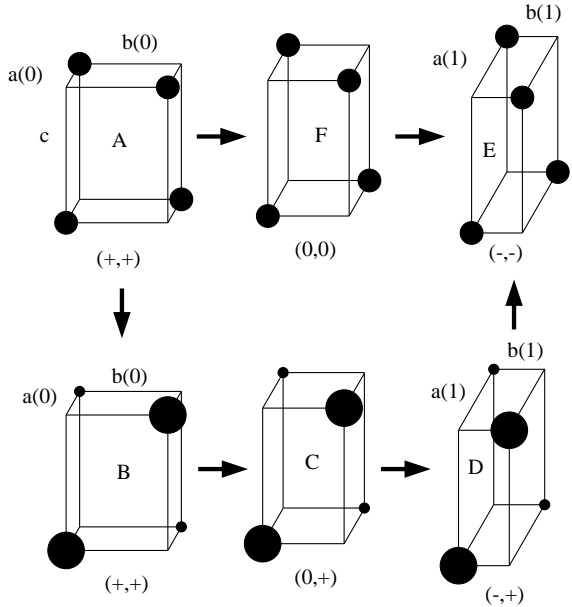}}
\vskip15pt
\caption{Path I (AFE) and path II (ABCDE) between
chiral enantiomers
(mirror images) A and E for a molecule consisting of $4$ atoms at
the vertices of a twisted H for the parameters used in Fig. \ref{chiralplot}.
Below each configuration we give
$(\sigma, \sigma')$, where $\sigma$ and $\sigma'$ are
respectively the signs or zero value of $\psi_0$ and
$\psi_0'$.  In configurations C and F, $a=b$.  Path $AFE$ passes
through the achiral configuration $F$.  Path $ABCDE$ passes
through chiral configurations only.  Configuration $C$ would
have a mirror plane (as does F) if the masses were all equal.}
\label {path}
\end{figure}

We will now consider a process in which the molecule is distorted
in the parameter space $(a,b,\mu)$
from the initial configuration ${\bf A}$ of Fig.\
\ref{path} into its enantiomer, ${\bf E}$, while
{\sl remaining chiral along the entire path of deformation}.
Initially $\mu=0$ and $0<a<b<c$.

\begin{figure}
\epsfxsize=2.5truein
\centerline{\epsfbox{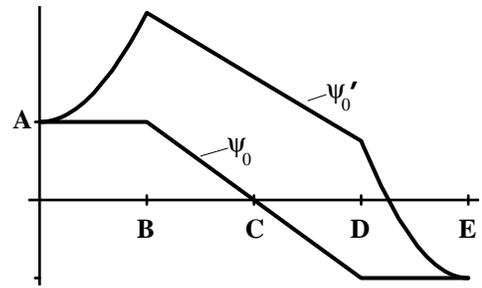}}
\vskip15pt
\caption{Plot of the chiral measure $\psi_0$ and
$\psi'_0$ along the path
${\bf ABCDE}$.  Note that both parameters pass through zero on this path, but
they do not pass through zero at the same place.  The parameters used to
obtain this plot are $a(0) = 0.99$, $b(0) = 1.01$, $c=1.2$, and $\mu_0 =
0.15$.}
\label{chiralplot}
\end{figure}

Note that any pseudoscalar must change sign under reflection
and therefore must pass through $0$ somewhere along the path between
enantiomers.  We will consider two paths between the molecule and its
mirror image described in Fig.\ \ref{path}:  the first will be a path $\bf AFE$
through an achiral point, the second a path $\bf ABCDE$ that goes only through
chiral states, along which
$\psi_0$ and $\psi_0'$ {\sl never} simultaneously vanish.  Our
deformation will rearrange the molecule into its mirror
image under the operation $(x,y,z)\rightarrow(y,x,z)$ which takes
$(a,b,c)$ into $(b,a,c)$.  We may parameterize $\bf AFE$ by
\begin{mathletters}
\label{ABEQ}
\begin{eqnarray}
a(t) &=& a(0) + [b(0)-a(0)]t \\
b(t) &=& b(0) + [a(0)-b(0)]t \\
\mu(t) & = & 0 \ ,
\end{eqnarray}
\end{mathletters}
where $t=0$ corresponds to the initial configuration ${\bf A}$
and $t=1$ to the enantiomer ${\bf E}$.  For concreteness, we have
used the values $a(0)=0.99$, $b(0)=1.01$ and $c=1.2$.
Note that $t=1/2$ corresponds to
the point ${\bf F}$, which is achiral, at which it is easy to see that
$\psi_0=\psi_0'=0$ since
$a(1/2)=b(1/2)$ and $\mu=0$.  To pass between enantiomers {\sl without}
passing through an achiral configuration we will follow the
path $\bf ABCDE$ along which the mass parameter $\mu$ does
not remain fixed at zero.
Note that this path avoids the line, $a=b$ and $\mu=0$, along which
the molecule is achiral.  Over the first section of the path,
${\bf AB}$, we change the masses of the atoms
by changing $\mu$ from its initial zero value to a suitable value
$\mu$ ($\mu_0=0.15$).
Over the second section, ${\bf BCD}$
the mass parameter is held fixed, so that $\mu=\mu_0$, but $a$ and $b$
are varied as in Eq. (\ref{ABEQ}a,b), so at ${\bf D}$ $\mu=\mu_0$, $a(t_D)=b$
and $b(t_D)=a$.
Finally along ${\bf DE}$,  $a(t)$ and $b(t)$ remain constant
but $\mu$ is changed from $\mu_0$ back to
zero.  For the parameter values we have chosen, $\psi_0'$ changes
sign on this part of the path.  Since
configuration ${\bf E}$ is the mirror image of configuration ${\bf A}$,
the values of all chiral parameters, including $\psi_0$
and $\psi_0'$ have changed sign.  But all states in the path of
deformation are chiral and nowhere on this path do both
$\psi_0$ and $\psi_0'$ simultaneously vanish!
The chiral measures $\psi_0$ and $\psi'_0$ for the path ${\bf ABCDE}$
are plotted in Fig.\ \ref{chiralplot}.

We note that in a molecule with more than four atoms, our artificial
deformation
of the masses can be replaced by the additional degrees of freedom provided
by the other atoms.

What shall we conclude from this example?
Since we can continuously deform a molecule into its
enantiomer via only chiral states, there is no general, unambiguous
characterization
of handedness.

\section{Prediction of the Cholesteric Pitch}

We now turn to the calculation of the cholesteric wavenumber $q_0$
in terms of the microscopic interactions between molecules.  We will argue
that previous classical analyses of this problem have missed an essential
feature
of chiral interactions -- the necessity of biaxial correlations between the
molecules.
Our result will involve the chiral parameter $\psi_2$ defined in
Eq.\ (\ref{psi2}).
In accordance with experimental observations we will assume that $q_0 a \ll 1$,
where $a$ is a typical intermolecular separation.  This means that the
cholesteric can be treated locally as a nematic even though
it is twisting on longer length scales.  Our aim, then, is to obtain
a formula for
$q_0$ in terms of correlation functions evaluated in the nematic limit, {\sl
i.e.}, when all chiral interactions have been turned off.  This type of result
is particularly desirable for numerical simulations, since it only
requires nematic correlation
functions and hence simulations of the nematic state
rather than the cholesteric state.

In this spirit we start by
considering the long-wavelength properties of systems that are
locally nematic with fluctuations described by the
phenomenological Frank free energy (Frank, 1958), which depends on the
director ${\bf n} ({\bf x})$.  When surface terms are neglected, which only
come into play when topological defects or internal surfaces are present
(Kl\'eman 1983), the Frank free energy is
\begin{eqnarray}
\label{FRANK}
F&=&{1\over 2}\int d^3\!x\,\Big\{K_1\left({\bf\nabla}\cdot{\bf n}\right)^2
+K_2\left({\bf n}\cdot{\bf\nabla}\times{\bf n} \right)^2\nonumber\\
&&\quad\quad +K_3
\left[{\bf n}\times\left({\bf\nabla}\times{\bf n}\right)\right]^2
+ 2h {\bf n} \cdot \nabla \times {\bf n} \Big\} .
\end{eqnarray}
This effective free energy can describe most of the phenomena of nematic and
cholesteric liquid crystals.  The parameter $h$ is the
generalized thermodynamic force that
determines the pitch.  We shall refer to $h$ as the torque field since
it is proportional to the microscopic intermolecular torques as we shall show.
To see how $h$ determines the pitch,
we consider a helical configuration of ${\bf n}( {\bf x})$
as in Eq.\ (\ref{twistn}). Then, the twist
\begin{equation}
- {\bf n} \cdot \nabla \times {\bf n} = q
\end{equation}
is spatially uniform, ${\bbox{\nabla}} \cdot {\bf n}=0$, and ${\bf n}
\times ({\bbox{\nabla}} \times {\bf n}) = 0$.  The Frank free energy is
\begin{equation}
\label{freesmall}
F= \Upsilon\left\{{1\over 2}K_2q^2 - hq \right\},
\end{equation}
where $\Upsilon$ is the volume of the system.  This energy is minimized when
\begin{equation}
q = q_0 = h/K_2 .
\label{q_0}
\end{equation}
Note that $h$ is a pseudoscalar.
Under spatial inversion, $h$ and,
therefore,
also the ``twist'' change sign: ${\bf n}\cdot{\bf\nabla}\times{\bf n}
\rightarrow -{\bf n}\cdot{\bf\nabla}\times{\bf n}$ since it
is linear in spatial gradients.  To obtain a nonzero value of $h$ it is,
therefore, necessary that the system {\sl not} be invariant under
spatial inversion.  Furthermore, we see that
to evaluate $q_0$ we need to calculate both $h$ and $K_2$.

The Frank elastic constants $K_1$, $K_2$, and $K_3$ can be estimated by
dimensional analysis using only excluded volume (entropic) interactions by
dividing an energy scale by a length scale.  Taking the energy scale
to be $k_{\scriptscriptstyle B}T\sim 4\times 10^{-14} \rm erg$ and the
length scale to be a molecular size $\sim 1 \,\rm nm$, we find that
$K_i\sim 4\times 10^{-7} \rm dyne$, or about $1\,\,\mu\,\rm dyne$.  This is
approximately correct for typical liquid crystals that
exist at room temperature.  As we discussed in the introduction, dimensional
analysis does not predict $q_0$ correctly: $q_0^{-1}$ is
typically on the order of or larger than hundreds of nanometers and does not
correspond to any natural length scale in the problem.  The challenge,
therefore, is to calculate $h$ and to determine why its magnitude does not
correspond to what dimensional analysis suggests.

A major objective is to establish an approach that in principle will
provide a rigorous calculation of the cholesteric pitch (or equivalently of
$h$) in the limit when the pitch is very long, {\sl i.e.},
when $q_0 a \rightarrow 0$.
The starting point of any microscopic calculation of
$h$ must be the intermolecular potentials.  Liquid crystal mesogens are
notoriously complex, containing hundreds to hundreds of thousands of individual
atoms, and they have correspondingly complex interactions. A reasonable
approach, and the one we will pursue here, to construct the desired
potentials is to model each mesogen as a collection of connected spherically
symmetric atoms that interact via pairwise central-force potentials with atoms
on other mesogens.  The interatomic potential consists of a long-range van der
Waals part and a short-range repulsive part arising mostly from the Pauli
exclusion principle.  Fluid physics is dominated by the short-range repulsive
part, and its is often useful to replace the full interatomic potential by a
simple hard-core potential with no attractive part.  An intermolecular
potential constructed in this way includes steric interactions that force two
chiral objects like screws or ridged ``barber poles'' (Fig.\
\ref{straley_screws}) to twist relative to each other when in contact.

There are contributions to the effective intermolecular potential that cannot
be expressed as a superposition of interatomic central-force potentials.  The
simplest such contribution is a chiral dispersion potential -- an anisotropic
generalization of the van der Waals potential.  Dispersion forces arise from
the Coulomb potential between all pairs of electronic and nuclear charges
and quantum fluctuations of the electronic states.  The van der Waals
potential is produced by the interaction of fluctuating electric dipoles
on different atoms.  If molecules are chiral, a fluctuating dipole on
one molecule can interact with a fluctuating quadrupole on another to
produce an effective chiral dispersion force (van der Meer
{\sl et al.}, 1976; Kats, 1978; Issaenko {\sl et al.}, 1998).
It is difficult at the moment to obtain first-principles
estimates of the strength of chiral dispersion forces.

In what follows, we confine our attention to pairwise central-force
interactions.
Once the intermolecular potentials have been chosen, the next step is to
devise a scheme to compute $h$. Since $q_0 a \ll 1$, one may assume that $h$
is small and calculate all quantities to lowest order in $h$, or equivalently
to lowest order in $q_0 a $.  The cholesteric twist induces biaxial
contributions to the nematic order parameter or order $(q_0 a )^2$ (Priest and
Lubensky, 1974).
Thus, to lowest order in $q_0 a$, biaxiality can be ignored, and the
cholesteric can be treated as though it were locally uniaxial. Mean-field
theory
is a natural first calculational approach to pursue
(Schr\"oder, 1979; Evans, 1992; Pelcovits, 1996; Moro, {\sl et. al.}, 1996).
In the locally uniaxial limit appropriate to most cholesterics, mean-field
theory will always predict $h=0$ (Salem {\sl el al.} 1987) when central-force
potentials between atoms are assumed.  This result is easy to understand:
Mean-field calculations seek the best self-consistently determined distribution
function for a single mesogen.  In a uniaxial system, this distribution
function will be uniaxial and produce only uniaxial average mass moments. Since
there are no uniaxial structures that are chiral, any manifestation of
chirality is washed out, there will be no potential favoring relative
twist of neighboring molecules, and $q_0$ will be zero. Thus, a more powerful
approach than mean-field theory is needed to calculate $q_0$ in the majority of
cholesterics that are nearly uniaxial.  In the less common case that would
arise when chirality is introduced in a biaxial nematic, the cholesteric is
locally biaxial, and mean-field theory will produce a nonvanishing value of
$q_0$.

The failure of mean-field theory can be traced to its neglect of
biaxial correlations between neighboring
molecules.  A first principles theory developed by the authors
(Harris {\sl et al.}, 1997)
provides a rigorous method, not limited to mean-field theory, for
calculating $h$.  Its principal result is that, under certain
approximations, $h$ is proportional to a measure $\psi$ of molecular chirality
times the spatial integral of a biaxial correlation function -- a function
which is strictly zero
in mean-field theory.  Thus, $h$ is small and deviates from expectations
based on dimensional analysis both because $\psi$ can be small
and because biaxial correlations may be very short-ranged.
Here we outline some important features of this theory.  It begins with a
rigorous expression of $h$, which can be obtained from Eq.\
(\ref{freesmall}):
\begin{equation}
\label{heq}
h = -{1\over \Upsilon}\left.{\partial F\over\partial q}\right\vert_{q=0}\ .
\end{equation}
In the Appendix we show that within certain simplifying conditions
this formulation leads to the result for the cholesteric wave vector,
\begin{equation}
\label{qzeroeq}
q_0 = - {1\over 4K_2\Upsilon}\Bigg\langle\sum_{\scriptscriptstyle BA}
{\bf R}_\perp\cdot \bbox{\tau}_{\scriptscriptstyle BA}\Bigg\rangle ,
\end{equation}
where $\bbox{\tau}_{\scriptscriptstyle BA}$ is the torque exerted on molecule
$B$ by molecule $A$:
\begin{equation}
\label{toexpand}
\tau^i_{\scriptscriptstyle BA} = \sum_{\beta\alpha} \epsilon_{ijk}
r_{{\scriptscriptstyle B}\beta}^j \partial_k V\left({\bf R}
+ {\bf r}_{{\scriptscriptstyle
B}\beta} - {\bf r}_{{\scriptscriptstyle A}\alpha}\right) ,
\end{equation}
and $\langle\cdot\rangle$ denotes thermodynamic averaging.
It is no surprise that the intermolecular torques are
the origin of the cholesteric structure.  Indeed, if we had
considered two planes of molecules a distance $L$ apart along an
axis perpendicular to the nematic director, then the change
in angle between them would be $\theta=qL$ and
thus the expression for
$h$ in Eq.\ (\ref{heq}) would become
\begin{equation}
\label{rough}
h = -{L\over\Upsilon}\left.{\partial
F\over\partial\theta}\right\vert_{\theta=0}\ ,
\end{equation}
which is simply the torque per unit area.
Moreover, Eq.\ (\ref{qzeroeq}) provides a rigorous small-$q_0$
expression for $q_0$
in a fully aligned nematic
in terms of quantities to be evaluated in the
nematic limit, {\sl i.e.}, when molecular chirality is turned off.
It is interesting to observe that this result does not involve
simply the torque $\bbox{\tau}_{\scriptscriptstyle A} \equiv
\sum_{\scriptscriptstyle B} \bbox{\tau}_{\scriptscriptstyle BA}$
on molecule A.  In the nematic phase the average torque on
molecules in the interior of the sample is zero. Whether or not
the nematic is locally unstable relative to states with nonzero
$q$ depends on the boundary conditions (Harris {\sl et al.},
1999).  Accordingly, Eq.\ (\ref{qzeroeq}) involves what we call the
``projected torque on molecule A,'' namely $\sum_{\scriptscriptstyle B}
{\bf R}_{\perp} \cdot
\bbox{\tau}_{\scriptscriptstyle BA}$.
Finally, the appearance of the antisymmetric tensor in Eq.\ (\ref{toexpand})
guarantees that $q_0$ is a pseudoscalar and hence must vanish for a
system in which all molecules are achiral.

We now discuss some of the implications of the result in
Eq.\ (\ref{qzeroeq}).  For that purpose we consider a number
of approximations that lead to a simple, yet nontrivial case.
First, the molecules were assumed to
have their long axes perfectly aligned along the director, {\sl i.e.} their
principal axes vectors ${\bf e}_3$ are parallel to ${\bf n}$.  Although
strictly speaking this limit is not realized in real systems, it does enable us
to see some simple consequences of our formalism. Second, we neglect
correlations between density fluctuations and orientational fluctuations.
Finally, we will invoke an expansion in powers of $r/R$.  Elsewhere
(Harris {\sl et al.}, 1999) we give a less restrictive discussion.
As our prior discussion indicates, we must be sure to take the uniaxial
average of the molecular
orientations.
On doing so, we found (Harris {\sl et al.}, 1997)
that the first non-zero term upon averaging was fifth order in
powers of $r$.  The lowest order expression for $q_0$ depends on
the nematic alignment tensor
$Q^{ij}$,
the biaxial tensor $B^{ij}$, and the third-rank tensor $S^{ijk}$ introduced
in Eq.\ \ref{SBQEQ}.  We found
\begin{equation}
\label{finalanswer}
q_0 = -{\sum_{\scriptscriptstyle BA}\epsilon_{ijk}
Q^{ip}\left\langle\left(B^{jl}_{\scriptscriptstyle
B}S^{kpl}_{\scriptscriptstyle A}
+B^{jl}_{\scriptscriptstyle A}S^{kpl}_{\scriptscriptstyle B}\right)K({\bf
R})\right\rangle\over 8K_2\Upsilon} \ ,
\end{equation}
where $K({\bf R})$ is a sum of products of $\vert{\bf R}\vert$ and
derivatives of the interaction potential $V({\bf R})$.

The sum in Eq. \ (\ref{finalanswer}) is averaged over molecular orientations
and locations.  The tensors $B^{ij}$ and $S^{ijk}$
depend on
the orientation of the molecule.  In fact, if the molecules
spin independently about their long axes,
this average of $B^{ij}_{\scriptscriptstyle X}$ will vanish.
Moreover, the components of
$S^{ijk}_{\scriptscriptstyle X}$ that contribute to Eq. \ (\ref{finalanswer}),
$\overline{S}^{ijk}$, can be expressed in terms of
$B^{ij}_{\scriptscriptstyle X}$
so that, for identical molecules, Eq. \ (\ref{finalanswer}) becomes
\begin{equation}
q_0 = - \psi_2{\sum_{\scriptscriptstyle BA}\left\langle
\left(B^{ij}_{\scriptscriptstyle B}B^{ij}_{\scriptscriptstyle A}
\right)K({\bf R}) \right\rangle\over 8K_2\Upsilon {\rm Tr}(B^2)} \ ,
\label{finalformof}
\end{equation}
where $\psi_2\equiv S^{klm}\epsilon_{ijk}Q^{il}B^{jm}$, as defined in
Section II, is
evaluated on a {\sl single molecule}.
Hence $\psi_2$ is a pseudoscalar parameter that is a measure
of the chiral interaction between identical chiral molecules.
It vanishes when the molecules are not chiral, and it
provides a quantitative
index
of chirality as would be measured through the cholesteric pitch.
We emphasize, however, that other microscopic measures of chirality will in
general involve other chiral parameters.
Note that since the biaxial correlations can be negative at the
intermolecular separation, there is not even a correlation between
the {\sl signs} of $\psi_2$ and the cholesteric pitch.

The correlation function in Eq.\ (\ref{finalformof}) may be evaluated in the
decoupling approximation where the molecular separation ${\bf R}$ is
uncorrelated with the biaxial orientation.  In this case the biaxial
correlation function is simply the average
\begin{equation}
\left\langle\left(B^{ij}_{\scriptscriptstyle B}B^{ij}_{\scriptscriptstyle A}
\right) \right\rangle =\left\langle\cos\left[2\left(\phi_{\scriptscriptstyle
A}-\phi_{\scriptscriptstyle B}\right)\right]\right\rangle,
\end{equation}
where
(see Fig.\ 1c)
$\phi_{\scriptscriptstyle A}$ is the angle between the biaxial
axis of molecule $A$ and the $x$-axis.

In the above calculation of $q_0$, we assumed that all molecules in the
cholesteric were chiral.  In order for a chiral interaction producing relative
twist between two molecules to exist, however, it is only necessary for one
molecule to be chiral.  The linearity of the our expression for $q_0$ [Eq.\
(\ref{finalformof})] in $\psi_2$ is a consequence of this fact.  If both
molecules had to be chiral, one might expect $q_0$ to be proportional to
$\psi_2^2$.  This is impossible, of course, because a pseudoscalar ($q_0$)
cannot be proportional to the square of a pseudoscalar ($\psi_2$).
The expression for $q_0$ for a system composed of a mixture of
chiral and achiral molecules is essentially the same as Eq.\
(\ref{finalformof}) except that $A$ and $B$ refer to different molecular
species  and $\psi_2$ is the chiral parameter of the chiral molecule.  There
are also chiral interactions between a chiral molecule and a strictly uniaxial
molecule.  These interactions lead to contributions to $q_0$ that depend on
correlations between the chiral parameter of the chiral molecule
and the biaxial anisotropy of the positional correlation of its uniaxial
neighbor.
Thus, when a nematic is doped with chiral molecules, a
finite pitch must result, and we expect $q_0$ to be proportional to the
dopant concentration when it is small.

\section{Discussion and Conclusion}

In this paper we considered ways of characterizing and quantifying
molecular chirality and of calculating the pitch wavenumber $q_0$, a
macroscopic manifestation of chirality in cholesteric liquid crystals.  We
showed that there is not one, but an infinite number of chiral parameters that
characterize a chiral object.  Each chiral parameter is a pseudoscalar, whose
construction, if it is obtained from mass or charge distributions, requires the
contraction of at least three
mass- or charge-moment tensors. Chiral parameters for
a given object can have varying magnitudes and even different signs.  It is
possible to pass continuously from a chiral object to its mirror image without
ever passing through a state in which the object is achiral.  In this process,
each chiral parameter passes through zero, but at no point do all parameters
pass simultaneously through zero. We showed in Section IV that the macroscopic
pitch depends on both molecular chiral parameters and on molecular
orientational
correlations.  Since these correlations vary with temperature, $pH$, pressure,
{\sl etc.},
it is possible to change the magnitude of the cholesteric pitch without
changing molecules.
This mechanism may be the explanation for the phenomena of twist inversion
(Stegemeyer {\sl et al.}, 1989) in which the pitch changes continuously
from right- to left-handed as a function of temperature.
Moreover, orientational order ({\sl e.g.} hexatic)
can selectively enhance different
intermolecular correlations and thus change the importance of different chiral
parameters for determining the macroscopic cholesteric pitch.

Finally, we emphasize the usefulness of the formal development of
Section IV. It represents an important advance in that in the
limit of long pitch, it gives an expression for the pitch in terms
of quantities in the nematic system when chiral interactions have
been turned off.  Especially for simulations, this implies that
it is not necessary to simulate a long pitch system.  Instead one
can simulate a homogeneous nematic in order to get correlation
functions of the type appearing in Eq.\ (33).  Further analysis
in this direction may be needed to actually implement this idea.

\section*{Acknowledgments}
It is a pleasure to acknowledge stimulating discussions with Joel Schnur
and David Walba.  ABH was supported by NSF Grant Number DMR95-20175.
RDK was supported through an NSF Career Award through Grant Number
DMR97-32963, an award from Research Corporation and a gift from L. J.
Bernstein.
TCL was supported by NSF Grant Number DMR97-30405.

\section*{Glossary of Terms}
For the non-specialist, we include the following glossary of technical terms:

\begin{enumerate}
\item {\bf cholesteric}: a material in the cholesteric (or
twisted nematic) phase. The director in this phase
has a helical structure (depticted in Fig.\ 3) obtained
obtained by twisting a nematic.
\item {\bf director}: unit vector ${\bf n}$ specifying the
direction of average orientation of anisotropic molecules in
a mesophase.
\item {\bf enantiomer}: a molecule with a given chemical formula
can can exhibit many different geometrical structures
called isomers.  A chiral isomer is an enantiomer.  A chiral
isomer and its mirror image are an enantiomeric pair.
\item {\bf Frank free energy}: energy [Eq.\ (30)] associated with
long-wavelength distortions of the director in a nematic.
It is proportional to the square of spatial derivatives of
$\bf n$.
\item {\bf mesogen}: a molecule that forms a mesophase.
\item {\bf mesophase}: a phase with symmetry intermediate between
that of the most disordered isotropic, spatially homogeneous
fluid phase and those of the most ordered period crystal
phases. All liquid-crystal phases except for those, such as
the cholesteric blue phase, that have true three-dimensional
periodic order are mesophases.
\item {\bf nematic}: a liquid crystalline material composed of
anisotropic (rod or disk shaped) particles with long-range
orientational but no long-range translational order. This term
comes from the Greek word
$\nu \epsilon \mu \omega \sigma$ for thread.  A nematic is
often filled with defects that look like threads under cross
polarizers.
\item {\bf nematogen}: a molecule that forms a nematic phase.
\item {\bf smectic}: from the Greek word
$\sigma\mu\epsilon\gamma\mu\alpha$ for soap. A smectic phase
is a ``solid" in one dimension and a fluid in the other two
directions. It consists of equally spaced parallel layers.
\item {\bf steric}: arising from hard-core, excluded-volume
interactions.  This term comes from the Greek word $\sigma\tau
\epsilon\rho\epsilon{\sl o}\sigma$ for solid.
\end{enumerate}

\begin{appendix}
\section{Expression for the Torque Field}
In this appendix we recast the expression for $h$ in a simplifying
limit, {\sl viz.} when the molecules are perfectly aligned along
the local nematic axis (but their biaxial axis is not fixed).
We start from Eq.\ (\ref{heq}).  As mentioned in the text, it
is clear that $h$ is zero if the system is achiral.  Thus
we are interested in the terms in the free energy which are linear
in the chiral parameters, $\psi_n$.  So we may write
\begin{eqnarray}
\label{APPEQ}
h = - \left. {1 \over \Upsilon} \sum_n \psi_n {\partial^2 F \over
\partial q \partial \psi_n} \right \vert_{q=0, \psi_n=0} \ .
\end{eqnarray}
Note that $q$ enters the calculation in the following way.
We consider a helical phase in which the director $\bf n$ is given
by Eq.\ (\ref{neq}).  In the small $q$ limit,
the atomic coordinates are displaced by an amount
$\delta r^i_{{\scriptscriptstyle A}\alpha}$ from their
reference positions in the nematic phase, where
$\delta r^i_{{\scriptscriptstyle A}\alpha}
= \epsilon^{ijk} \delta\omega^{j}_{\scriptscriptstyle A}
r^k_{{\scriptscriptstyle A}\alpha}$,
$\delta\omega^j_{\scriptscriptstyle A}
= q e^j\left({\bf e}\cdot{\bf R}_{\scriptscriptstyle A}\right)$ and
where ${\bf e}$ is an arbitrary unit vector perpendicular to ${\bf n}$.
In this sense the total potential energy $U$ has a $q$-dependence
such that
\begin{eqnarray}
\label{A2}
{\partial U \over \partial q} & = &
\sum_{{\scriptscriptstyle A}\alpha i} {\partial U \over
\partial r^i_{{\scriptscriptstyle A}\alpha}}
{\partial r^i_{{\scriptscriptstyle A}\alpha} \over \partial q} \ .
\end{eqnarray}
With this understanding, one evaluates Eq. (\ref{APPEQ}) as
\begin{eqnarray}
\label{FULLHEQ}
h &=& - {1\over\Upsilon} \sum_n \psi_n \left[ \left \langle
{ \partial ^2 U \over \partial q \partial \psi_n}
\right \rangle + \chi_n \right] \ ,
\end{eqnarray}
where $\langle\cdot\rangle$ denotes a thermodynamic average in
which the density matrix $\exp [- U/(k_{\scriptscriptstyle B}T)]$
is evaluated when
all molecular chirality is turned off and
\begin{eqnarray}
\chi_n & = & {1\over k_{\scriptscriptstyle B}T}\left \langle\,
{\partial U \over \partial q}
{\partial U \over \partial \psi_n }\, \right \rangle \ .
\end{eqnarray}

Note the appearance in $h$ of the terms in $\chi_n$.  As we will
show elsewhere (Harris, {\sl et al.} 1999), these terms, which normally are
not considered, are needed to obtain the expected result that
$h$ vanishes in the limit of an isotropic fluid (for
which the nematic order parameter vanishes).  However, in the
limit of nearly complete nematic order (which we consider here),
these terms in $\chi_n$ are negligible.  Superficially it may
appear that we have to
isolate the dependence of $U$ on the chiral parameters.  However,
since achiral components of $U$ do not contribute to
$\langle\, \partial U / \partial q \,\rangle$, this step is,
in fact, not necessary, so that
\begin{eqnarray}
\label{A5}
h = - {1\over\Upsilon}  \left \langle\,
{\partial U \over \partial q }\,\right \rangle \ .
\end{eqnarray}
This equality is the basis of our calculation -- it allows us to
calculate $\partial F/\partial q$ {\sl microscopically} in terms of
molecular interactions.  In particular, note that the result is
expressed in terms of a correlation function to be evaluated
in the nematic ($q=0$) limit (which we do implicitly in the following.)
It is important to note that in the more realistic limit when
the orientations of the molecules fluctuate away from the local
nematic direction, both terms in Eq. (\ref{FULLHEQ}) must be retained.
Such a calculation has not yet been carried out.

Writing the potential energy as a sum of identical (for simplicity),
pairwise central-force interactions $V({\bf R})$ and using
Eqs.\ (\ref{q_0}), (\ref{A2}), and (\ref{A5}), we find
\begin{eqnarray}
\label{thetorque}
q_0 & = & -{1\over 2K_2\Upsilon}\Bigg\langle\sum_{{\scriptscriptstyle
BA}\beta\alpha}
\epsilon_{ijk}
\partial_i V\left({\bf R}_{\scriptscriptstyle B} +{\bf r}_{{\scriptscriptstyle
B}\beta}
- {\bf R}_{\scriptscriptstyle A} - {\bf
r}_{{\scriptscriptstyle A}\alpha}\right)\nonumber\\
&&\quad\quad e_j\left\{\left({\bf e}\cdot{\bf R}_{\scriptscriptstyle
B}\right)r^k_{{\scriptscriptstyle B}\beta}- \left({\bf e}\cdot{\bf R}
_{\scriptscriptstyle A}\right)r^k_{{\scriptscriptstyle
A}\alpha}\right\}\Bigg\rangle \ .
\end{eqnarray}
Because the system is uniaxial, we can average over all perpendicular
directions $\bf e$, so that
$e_ie_j \rightarrow {1\over 2}(\delta_{ij} - n_in_j)$.  In this
case Eq.\ (\ref{thetorque}) becomes
\begin{eqnarray}
\label{torque}
q_0 & = & -{1\over 4K_2\Upsilon}\Bigg\langle\sum_{{\scriptscriptstyle
BA}\beta\alpha}
\epsilon_{ijk}
\partial_i V\left({\bf R} +{\bf r}_{{\scriptscriptstyle B}\beta} - {\bf r}_{
{\scriptscriptstyle A}\alpha}\right)\nonumber\\
&&\quad\quad\left[ R_\perp^j r^k_{{\scriptscriptstyle B}\beta} + R^j_{\perp
{\scriptscriptstyle B}} \left(r^k_{{\scriptscriptstyle B}\beta} -
r^k_{{\scriptscriptstyle A}\alpha}\right)\right]
\Bigg\rangle,
\end{eqnarray}
where ${\bf R}_{\perp}$ is the projection of $\bf R$ onto the plane
perpendicular to $\bf n$ and ${\bf R}\equiv
{\bf R}_{\scriptscriptstyle B} - {\bf R}_{\scriptscriptstyle A}$.  By
translational invariance, the second of the two terms inside the square
brackets of Eq.\ (\ref{torque}) must vanish:
for fixed molecule $B$, one can shift the origin of the coordinate
system by a fixed vector $\bbox{\Delta}$.  This would make the second term
depend on the choice of origin, which it cannot.  Hence we find that
\begin{equation}
q_0 = - {1\over 4K_2\Upsilon}\Bigg\langle\sum_{\scriptscriptstyle BA}
{\bf R}_\perp\cdot \bbox{\tau}_{\scriptscriptstyle BA}\Bigg\rangle ,
\end{equation}
where $\bbox{\tau}_{\scriptscriptstyle BA}$ is the torque exerted on molecule
$B$ by molecule $A$:
\begin{equation}
\tau^i_{\scriptscriptstyle BA} = \sum_{\beta\alpha} \epsilon_{ijk}
r_{{\scriptscriptstyle B}\beta}^j \partial_k V\left({\bf R}
+ {\bf r}_{{\scriptscriptstyle
B}\beta} - {\bf r}_{{\scriptscriptstyle A}\alpha}\right) .
\end{equation}
\end{appendix}

\end{document}